\documentclass[]{spie}  

\usepackage{amsmath,amsfonts,amssymb}
\usepackage{graphicx}
\usepackage[colorlinks=true, allcolors=blue]{hyperref}

\title{More chemical detection through less sampling: amplifying chemical signals in hyperspectral data cubes through compressive sensing}

\author[a]{Henry Kvinge}
\author[a]{Elin Farnell}
\author[b]{Julia R. Dupuis}
\author[a]{Michael Kirby}
\author[a]{Chris Peterson}
\author[b]{Elizabeth C. Schundler}
\affil[a]{Colorado State University, Department of Mathematics, 1874 Campus Delivery, Fort Collins, CO 80523-1874, USA}
\affil[b]{Physical Sciences Inc., 20 New England Business Center, Andover, MA 01810-1077, USA}

\authorinfo{Further author information (E-mail): (Send correspondence to H.K.)\\H.K.: henry.kvinge@colostate.edu; E.F.: elinfarnell@gmail.com; J.R.D.: jdupuis@psicorp.com;\\ M.K.: kirby@math.colostate.edu; C.P.: peterson@math.colostate.edu; E.C.S.: eschundler@psicorp.com}

\begin{document} 
\maketitle

\begin{abstract}

Compressive sensing (CS) is a method of sampling which permits some classes of signals to be reconstructed with high accuracy even when they were under-sampled. In this paper we explore a phenomenon in which bandwise CS sampling of a hyperspectral data cube followed by reconstruction can actually result in amplification of chemical signals contained in the cube.  
Perhaps most surprisingly, chemical signal amplification generally seems to increase as the level of sampling decreases. In some examples, the chemical signal is significantly stronger in a data cube reconstructed from 10\% CS sampling than it is in the raw, 100\% sampled data cube. We explore this phenomenon in two real-world datasets including the Physical Sciences Inc. Fabry-P\'{e}rot interferometer sensor multispectral dataset and the Johns Hopkins Applied Physics Lab FTIR-based longwave infrared sensor hyperspectral dataset. Each of these datasets contains the release of a chemical simulant, such as glacial acetic acid, triethyl phospate, and sulfur hexafluoride, and in all cases we use the adaptive coherence estimator (ACE) to detect a target signal in the hyperspectral data cube. We end the paper by suggesting some theoretical justifications for why chemical signals would be amplified in CS sampled and reconstructed hyperspectral data cubes and discuss some practical implications.

\end{abstract}


\section{INTRODUCTION}
\label{sec:intro}  

One of the most important applications of hyperspectral imaging is to the problem of detecting specific chemicals in a given scene. Unfortunately, hyperspectral images are generally much larger than traditional images even when one is collecting relatively few bands, and hyperspectral devices also generally come with a high price tag. For these reasons, any methods that allow hyperspectral images to be sampled at lower rates while at the same time retaining their discriminative ability have many applications. In particular, strategies for reduction of the number of sensors are quite valuable. Compressive sensing (CS) is exactly such a framework. In particular, CS provides methods for accurately reconstructing under-sampled signals.

Applying CS to hyperspectral imaging is an active area of research, with a multitude of different approaches. These include CS sampling schemes that are performed to different extents in both the spatial and the spectral domain. Several works have gone beyond simply trying to reconstruct the image itself and instead make the process of signal unmixing, i.e. identifying the different spectral signals in an image, a component of the reconstruction process \cite{martin2012new,martin2013hyperspectral,martin2015hyca}. Other works have explored reconstruction frameworks specifically designed to utilize the structure of hyperspectral data \cite{golbabaee2012hyperspectral}. Finally, various CS strategies have been proposed for data extracted from devices specific to hyperspectral imaging \cite{rajwade2013coded}.

In this paper we explore chemical detection in CS reconstructions of hyperspectral images after low levels of sampling. We show that, surprisingly, not only can a chemical still be detected in a hyperspectral image that has been reconstructed after 10\% sampling, but at least in some examples, the chemical signature can be slightly stronger in the reconstructed image. This is a surprising result because naively, one would expect that by using 10\% sampling, one is losing 90\% of the information from the hyperspectral image. We note that signal amplification only seems to happen with low sampling levels. An interesting future direction would be to understand in which situations precisely a signal is amplified by reconstruction.

This paper is organized as follows. In Sec.~\ref{sec:compressive_senseing}, we give a brief summary of some of the ideas underlying CS. In Sec.~\ref{sec:Framework}, we set up some of the mathematical framework and notation for discussing hyperspectral imagery. In Sec.~\ref{sect:dection_algorithms}, we describe the chemical detection algorithm (ACE) used in experiments for this paper. The main results of the paper are discussed in Sec.~\ref{sec:signa_enhancement}. Finally, in Sec.~\ref{sect:explainations}, we suggest some possible explanations for the chemical signal amplification seen in reconstructed images. In Sec. \ref{sec:conclusion} we suggest some directions for future research.

\section{BACKGROUND}

In this paper we will generally use upper case letters (for example $S$, $X$, $U$, $H$, and $Y$) to denote matrices. We will use lower case letters (for example $u$, $u^*$, $x'$, $\tilde{x}$, and $y$) to denote vectors. Matrices being placed side by side always denotes standard matrix multiplication.

\subsection{Compressive sensing and the single-pixel camera framework}
\label{sec:compressive_senseing}

Compressive sensing (CS) is a collection of methods that permit highly accurate reconstruction of certain classes of signals even when they have been sampled well under Nyquist-rate~\cite{baraniuk2007compressive}. When it is applicable, CS allows one to solve the ill-posed problem of finding $\tilde{x}$ from 
\begin{equation} \label{eqn-basic-CS-problem}
y = S\tilde{x} \in \mathbb{R}^k
\end{equation}
when $S$ is a $k \times n$ matrix, $\tilde{x} \in \mathbb{R}^n$, and $k < n$. A basic requirement for application of methods from CS for solution of \eqref{eqn-basic-CS-problem} is that we can make some assumption about the signal $\tilde{x}$. This allows us to choose $\tilde{x}$ (or a close approximation to $\tilde{x}$) out of the infinite number of solutions $x'$ to $Sx' = y$. One common choice of assumption about $\tilde{x}$ that is used frequently in CS is that it will be sparse in some particular basis. Although when $\tilde{x}$ is an image it will almost never be sparse in its natural basis, it is generally true that it will be compressible (that is, approximately sparse) in a wavelet basis. 

Let $H, H^{-1}: \mathbb{R}^n \rightarrow \mathbb{R}^n$ be the $1$-dimensional Haar wavelet transformation and its inverse respectively. Then the CS reformulation of \eqref{eqn-basic-CS-problem} is to solve:
\begin{equation} \label{eqn-opt-problem}
    u^* = \underset{u \in \mathbb{R}^{n}}{\text{argmin}}\; ||u||_{\ell_1} \quad\quad \text{such that } \quad y = SH^{-1}u,.
\end{equation}
In words, this optimization problem seeks $x$ satisfying $Sx = y$ such that $x$ is maximally sparse in the $1$-dimensional Haar wavelet basis, that is, $||u||_{\ell_1}$ is minimized for $u = Hx$.  

It is important to note that even though the solution $x^* = H^{-1}u^*$ to \eqref{eqn-opt-problem} may often be a very good approximation to $\tilde{x}$, we always expect some information to be lost in the process of sampling $\tilde{x}$ with $S,$ in particular when $k \ll n$. In some cases the information lost in this process is mostly noise, leading to a reconstruction $x^*$ that is preferable to the original signal $\tilde{x}$. For example, another variant of CS reconstruction which minimizes total variation \cite{rudin1992nonlinear} is closely related to a family of denoising algorithms called total variation denoising. Such an observation may form the first step in explaining the phenomenon described in this paper.

\subsection{Framework and notation for the compressive sensing of hyperspectral data}
\label{sec:Framework}

Hyperspectral data is highly structured and it is essential to capture this structure when performing optimization routines such as \eqref{eqn-opt-problem}. In this paper we assume that a single band of our hyperspectral data has size $n$ (that is, as a 2-dimensional array it has size $n_1 \times n_2,$ where $n=n_1n_2$) and that there are $b$ bands. Given the above conventions, we will realize an $n_1 \times n_2 \times b$ hyperspectral data cube as an $n \times b$ matrix $X \in \mathbb{R}^{n \times b}$, where we have flattened each band to a column vector in $X$. Then, $U = HX$ is an $n \times b$ matrix where the columns of $U$ are the columns of $X$ transformed into the $1$-dimensional Haar wavelet basis. Our optimization problem \eqref{eqn-opt-problem} applied to a hyperspectral image is thus
\begin{equation} \label{eqn-hyper-opt-problem}
        U^* = \underset{U \in \mathbb{R}^{n\times b}}{\text{argmin}}\; ||U||_{\ell_1} \quad\quad \text{such that } \quad Y = SH^{-1}U,
\end{equation}
where $S$ is again a sampling matrix in $\mathbb{R}^{k\times n}.$ The $\ell_1$-norm above is applied to $U$ in the same way as it would be to $U$ flattened to a length $nb$ vector. Note that the expression $H^{-1}U$ is equivalent to taking the inverse of the 1-dimensional Haar wavelet transform for every band in the data cube flattened to a vector. We call the $k \times b$ output $Y = SX$ a $(100\frac{k}{n})$\% sampling of $X$. In this paper we will generally sample at $10\%$, so that we are effectively throwing out $90\%$ of the information in a data cube. It is for this reason that it is surprising that chemical signals in $X$ sometimes become stronger.

There has been considerable research toward developing feasible optimization problems that reconstruct the bands of a hyperspectral image non-independently \cite{golbabaee2012hyperspectral}. Such methods make use of the correlation between bands. We chose to study \eqref{eqn-hyper-opt-problem} because in many cases hardware and sampling constraints force bands to be reconstructed independently. It would be interesting to understand if solving more hyperspectral specific optimization problems also produce reconstructions with signal enhancement.

In all the experiments described in this paper, we constructed $S$ via a modified Walsh-Hadamard matrix \cite{farnell2019sampling}. We suggest investigation into whether other sampling methods also result in signal enhancement at low sampling levels.

There are a large number of algorithms that have been developed for solving optimization problems such as \eqref{eqn-opt-problem} and \eqref{eqn-hyper-opt-problem}. We choose to utilize the split Bregman method \cite{GO09} since it is fast, lightweight, and gives reliable convergence.

\subsection{Detection algorithms}
\label{sect:dection_algorithms}

Since the point of our experiments was to demonstrate a phenomenon in which compressive sensing followed by reconstruction of a hyperspectral image amplifies chemical signals, choosing the appropriate chemical detection algorithm was a key component of our work. 

In our experiments we use the \emph{adaptive coherence/cosine estimator} (ACE), a well-known technique used for chemical detection~\cite{scharf1996adaptive,kraut2001adaptive}. Let $s$ be a target spectral signature and let $x$ be a spectral signature in a specific pixel within a hyperspectral cube (in the literature, $x$ is the \emph{pixel under test} (PUT)). The ACE statistic is the square of the cosine of the angle between $s$ and $x$ relative to the background. To be precise, the ACE statistic is calculated as 
$$\frac{(s^T\Gamma^{-1}x)^2}{(s^T\Gamma^{-1}s)(x^T\Gamma^{-1}x)},$$
where $\Gamma$ is the maximum likelihood estimator for the covariance matrix of background data.

In addition to using ACE for chemical detection, we compute the \emph{bulk coherence (multipulse coherence) estimator}~\cite{pakrooh2017adaptive,pakrooh2017adaptiveb,scharf2017multipulse}. The bulk coherence statistic enhances the signal in neighborhoods that have several pixels with relatively high ACE values, a property that is appropriate for chemical release settings. If $c_i$ is the ACE statistic for pixel $i$ and a neighborhood of pixels is indexed by $i=1,\ldots,M,$ then the bulk coherence statistic is $$1-\prod_{i=1}^M(1-c_i).$$ ACE values near one result in a product whose terms are close to zero, resulting in a high bulk coherence value (near one). Experimentally, the bulk coherence statistic leads to improved chemical detection. We also add one additional filter that we refer to as \emph{persistence} in some cases: we set the value associated to a pixel to zero if its bulk coherence value doesn't stay above a pre-specified threshold for at least five consecutive time steps.

To facilitate objective comparison, we further incorporate an algorithmic definition of a threshold for the ACE statistic (similarly for bulk coherence)~\cite{farnell2019TVvsL1}. We can then declare a chemical to be present or absent in a given pixel based on comparison of the ACE statistic against this threshold. The threshold definition is motivated by the idea that the threshold should be slightly larger than the ACE values observed in typical background cubes (cubes that have been collected with the intended device in which it is known that the target chemical is absent). The algorithmic determination of the threshold is responsive to the device used to sense the data, the reconstruction method, and the spectral signature of the target chemical.

\subsection{Signal enhancement in chemical detection}
\label{sec:signa_enhancement}

\begin{figure}[ht]
    \centering
    \includegraphics[width=14cm]{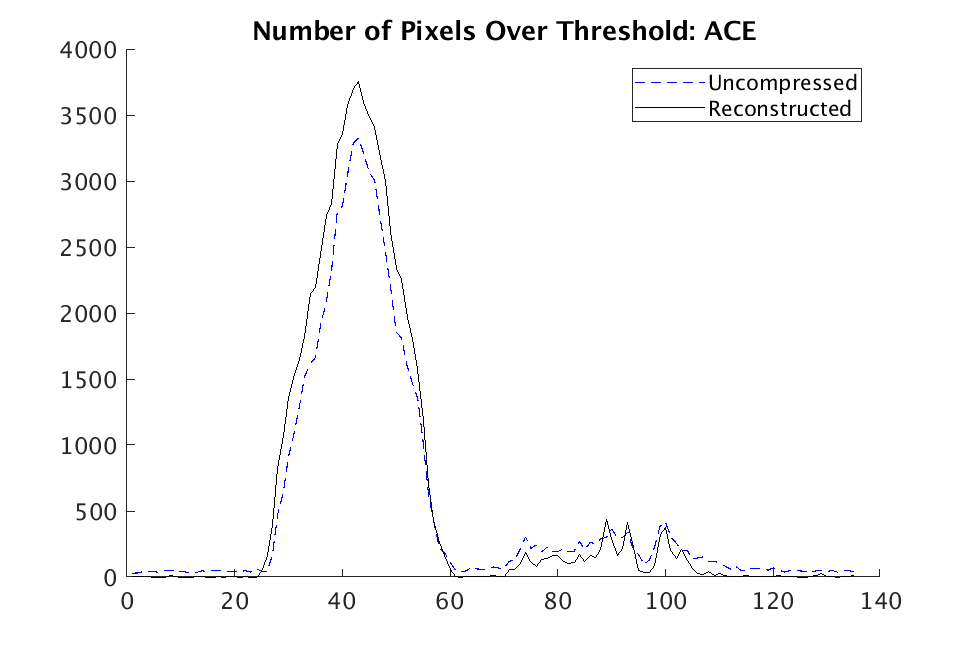}
    \caption{Comparison of chemical detection (Johns Hopkins SF6 27 Romeo dataset): the number of pixels which the ACE algorithm shows to contain the chemical signature for SF6 as a function of video frame for both uncompressed (dashed blue line) and compressively sensed and reconstructed data (solid black line). The $x$-axis is the frame number in the video while the $y$-axis is the number of pixels above the corresponding ACE threshold (where the threshold is as described in Sec.~\ref{sect:dection_algorithms}). During the peak of the release (around frame $40$), the signal is actually stronger in the hyperspectral cube that has been reconstructed from $10\%$ sampling.}
    \label{fig_sf6_number_over}
\end{figure}

\begin{figure}[ht]
    \centering
    \includegraphics[width=14cm]{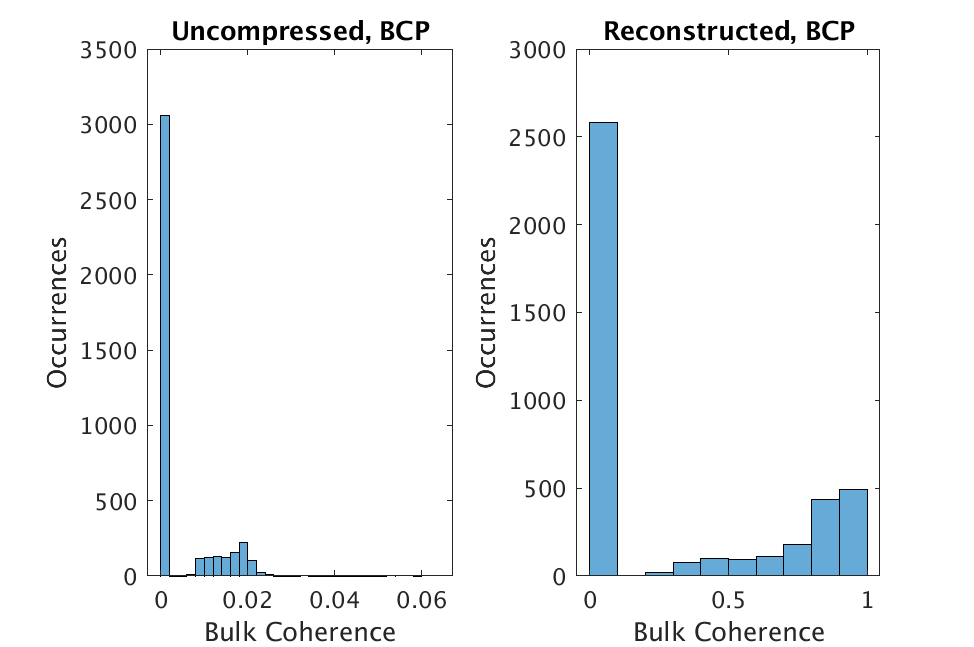}
    \caption{A histogram of ACE bulk coherence values with persistence for hyperspectral image $30$ from Fig.~\ref{fig_sf6_number_over}. The $x$-axis is the ACE bulk coherence value (larger values indicate a higher likelihood of the target chemical being present in the pixel). The bin close to zero consists of pixels not containing the chemical whereas the cluster of bins to the right is the collection of pixels that contain the chemical. As can be seen, there is a much wider spread between these two classes in the reconstructed image compared to what is found in the raw image.}
    \label{fig-SF6_Cube30_Hist}
\end{figure}

\begin{figure}[ht]
    \centering
    \includegraphics[width=14cm]{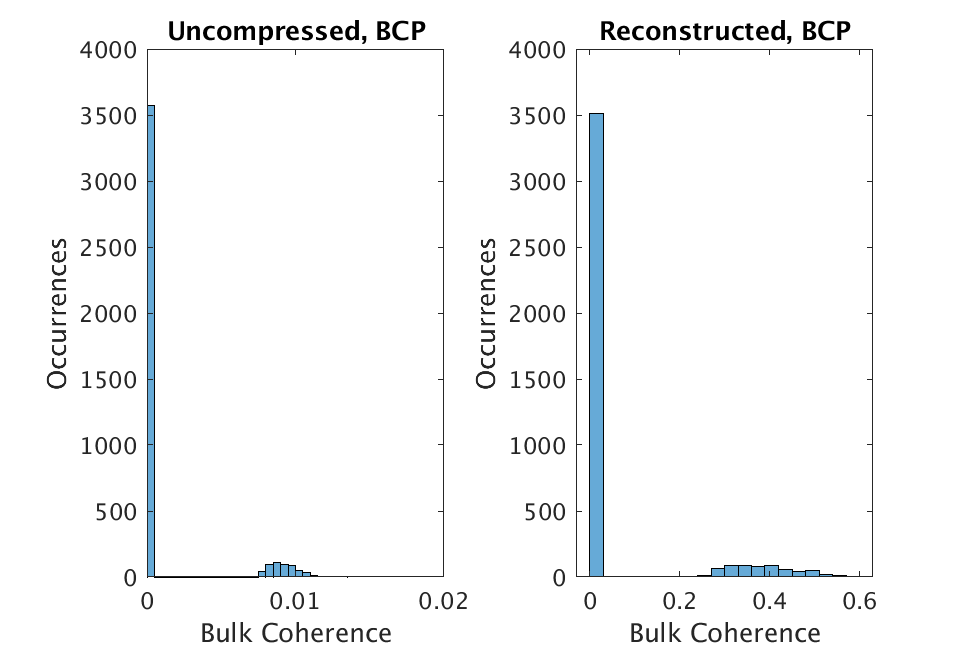}
    \caption{A histogram of ACE bulk coherence values with persistence for hyperspectral image $90$ from Fig.~\ref{fig_sf6_number_over}. The chemical has mostly dispersed from the scene at this point in the video. The $x$-axis is the ACE bulk coherence value (larger values indicate a higher likelihood of chemical being present in the pixel). The bin close to zero consists of pixels not containing the chemical whereas the cluster of bins to the right is the collection of pixels that contain the chemical. As can be seen, there is a much wider spread between these two classes in the reconstructed image compared to what is found in the raw image.}
    \label{fig-SF6_Cube90_Hist}
\end{figure}

We demonstrate the signal enhancement phenomenon on two hyperspectral datasets. The first is the Fabry-P\'{e}rot interferometer sensor multispectral dataset\cite{cosofret2009airis} and the second is the Johns Hopkins Applied Physics Lab FTIR-based longwave infrared sensor hyperspectral dataset~\cite{broadwater2011primer}. In our analysis of each data set, we restrict to a $64\times 64$ spatial field of view in which the chemical release is observed. Figures \ref{fig_sf6_number_over}-\ref{fig-SF6_Cube90_Hist} relate to the SF6 27 Romeo release video from the Johns Hopkins dataset. This video consists of $140$ hyperspectral images. Between image $20$ and $30,$ sulfur hexafluoride (SF6) gas is released and disperses. This release is apparent in Fig.~\ref{fig_sf6_number_over}, which shows the number of pixels that the ACE algorithm indicates contain the chemical as a function of time (the number of such pixels in the raw data is indicated by the dashed line and the number of pixels in the data reconstructed from $10\%$ sampling is indicated by the solid black line). It can be seen that during the release itself, more pixels in the reconstructed data have ACE values above the threshold than do those in the raw data. The chemical returns to the scene in a dissipated form sometime around cube 70 (presumably due to a change in wind direction). In this case, the uncompressed data results in slightly stronger chemical detection, with a few exceptions for short time frames. 

The specific distributions of bulk coherence ACE values with persistence for hyperspectral image $30$ (both raw (left) and reconstructed (right)), is shown in Fig.~\ref{fig-SF6_Cube30_Hist}. As this histogram indicates, the spread between values corresponding to spatial locations containing the chemical signature and those that do not is much larger in the reconstructed image. This is useful as it makes distinguishing pixels that do and do not contain the chemical easier. For reference the threshold value for chemical presence in the raw image is $0.0077$ and the threshold value for chemical presence in the reconstructed image is $0.2656$. 
A similar phenomenon is seen in Fig.~\ref{fig-SF6_Cube90_Hist} but for cube $90$ where most (but not all) the chemical has already dispersed from the scene.

\begin{figure}[ht]
    \centering
    \includegraphics[width=14cm]{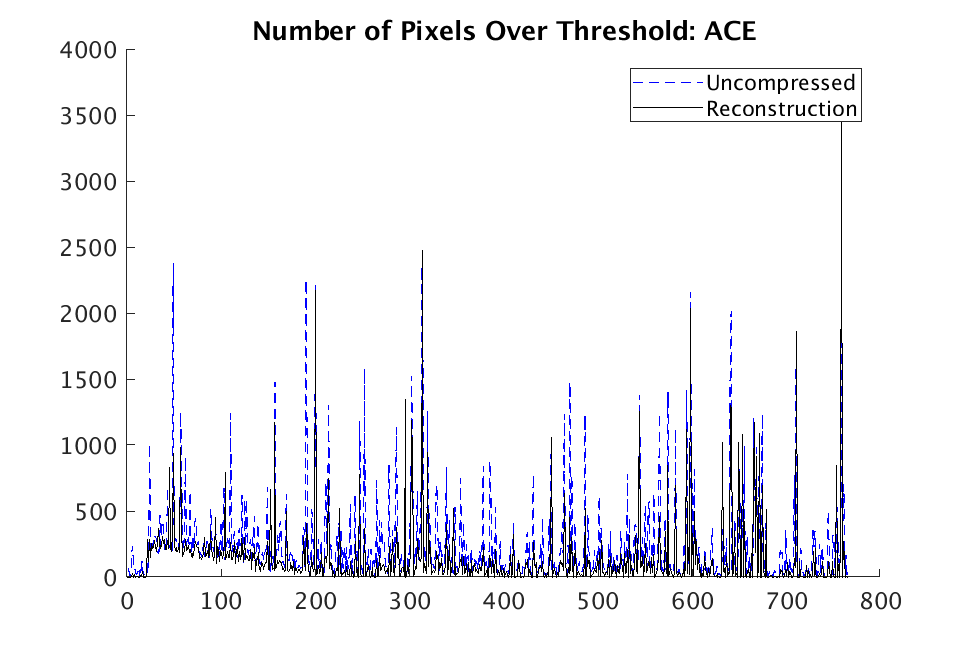}
    \caption{Comparison of chemical detection (Fabry-P\'{e}rot MeS C dataset): the number of pixels which the ACE algorithm shows to contain the chemical signature for MeS as a function of video frame for both uncompressed (dashed blue line) and compressively sensed and reconstructed data (solid black line). The $x$-axis is the frame number in the video while the $y$-axis is the number of pixels above the corresponding ACE threshold (where the threshold is as described in Sec.~\ref{sect:dection_algorithms}). Frequently, the signal in the reconstructed image is just as strong as the signal in the raw image despite the fact that this sequence of hyperspectral images has been reconstructed from $10\%$ sampling.}
    \label{fig-MESC_numberover}
\end{figure}

We next examine two hyperspectral videos in the Fabry-P\'{e}rot interferometer sensor multispectral dataset. The first contains a release of the chemical methyl salicylate (MeS). This release is shown in Fig.~\ref{fig-MESC_numberover}. As before the $x$-axis is the frame in the hyperspectral video and the solid black and dashed blue curve give the number of pixels that ACE indicates contain the chemical for reconstructed and uncompressed data, respectively. As can be seen, this is noisy data with many false positives (which appear as spikes). Despite the fact that the chemical signature in this dataset is quite weak even at the peak of the release, the signal has approximately the same strength in the reconstructed image compared to the raw image.

\begin{figure}[ht]
    \centering
    \includegraphics[width=14cm]{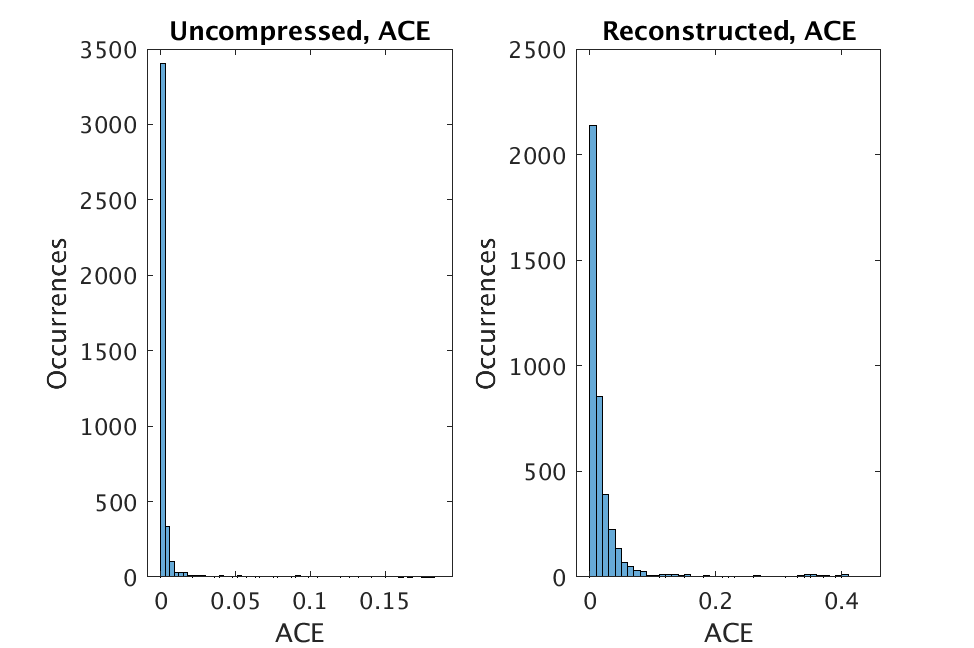}
    \caption{A historgram of ACE values for hyperspectral image $80$ in the MeS release (see Fig.~\ref{fig-MESC_numberover}). MeS gas is still present in the scene for this frame. The $x$-axis is the ACE value (larger values indicate a higher likelihood of chemical presence in the pixel). As can be seen, the reconstructed data (right) has a larger range of ACE bulk coherence values than the raw data. While spatial locations do not clearly separate into two classes, a wider spread in the data likely makes classification easier.}
    \label{fig-MESC_Hist}
\end{figure}

Figure \ref{fig-MESC_Hist} gives a histogram of ACE bulk coherence values for the MeS release at image 80 when the chemical is present in the scene. Unlike what was seen in Figs.~\ref{fig-SF6_Cube30_Hist} and~\ref{fig-SF6_Cube90_Hist}, there is no longer any clear separation between spatial locations containing the signature for MeS and spatial locations that do not. The fact that there is greater spread of values in the reconstructed image however makes classification of regions containing the chemical easier and more stable.

\begin{figure}[ht]
    \centering
    \includegraphics[width=14cm]{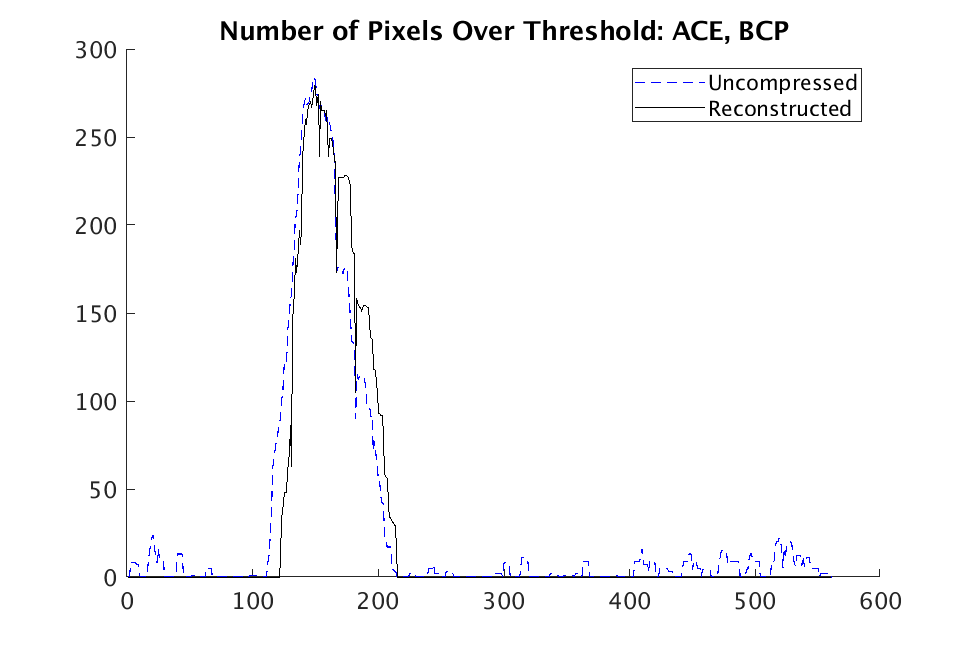}
    \caption{Comparison of chemical detection (Fabry-P\'{e}rot TEP A dataset): the number of pixels which the ACE algorithm shows to contain the chemical signature for TEP as a function of video frame for both uncompressed (dashed blue line) and compressively sensed and reconstructed data (solid black line). The $x$-axis is the frame number in the video while the $y$-axis is the number of pixels above the corresponding ACE threshold (where the threshold is as described in Sec.~\ref{sect:dection_algorithms}). After the peak of the release (around frame $160$), the signal in the reconstructed data is stronger than the signal in the raw data despite the fact that this hyperspectral data has been reconstructed from $10\%$ sampling. What is more, the uncompressed data exhibits noise throughout whereas the reconstructed data shows a strong signal when the chemical is present in the scene and does not show chemical present outside of that range.}
    \label{fig-TEPA_numberover_BCP}
\end{figure}

Finally, the Fabry-P\'{e}rot interferometer sensor multispectral dataset also contains a hyperspectral video of release of triethyl phosphate (TEP). A plot of this chemical release as captured by ACE is shown in Fig.~\ref{fig-TEPA_numberover_BCP}. This is an example where results of detection in the raw and reconstructed images are mixed. While the signal is better detected in raw data up to the peak of the release, it appears to be the case that the reconstructed data results in better detection in later frames as the chemical dissipates. 
This illustrates the point that signal amplification is not always consistent. Examples exist in which the signal is either slightly weaker or of equal strength in reconstructed images. Another important observation (which is discussed in Sec.~\ref{sect:explainations}) is that the false positives that appear in raw images are effectively eliminated in reconstructed images, pointing toward the reconstruction algorithm functioning as a denoising algorithm.

\begin{figure}[ht]
    \centering
    \includegraphics[width=10cm]{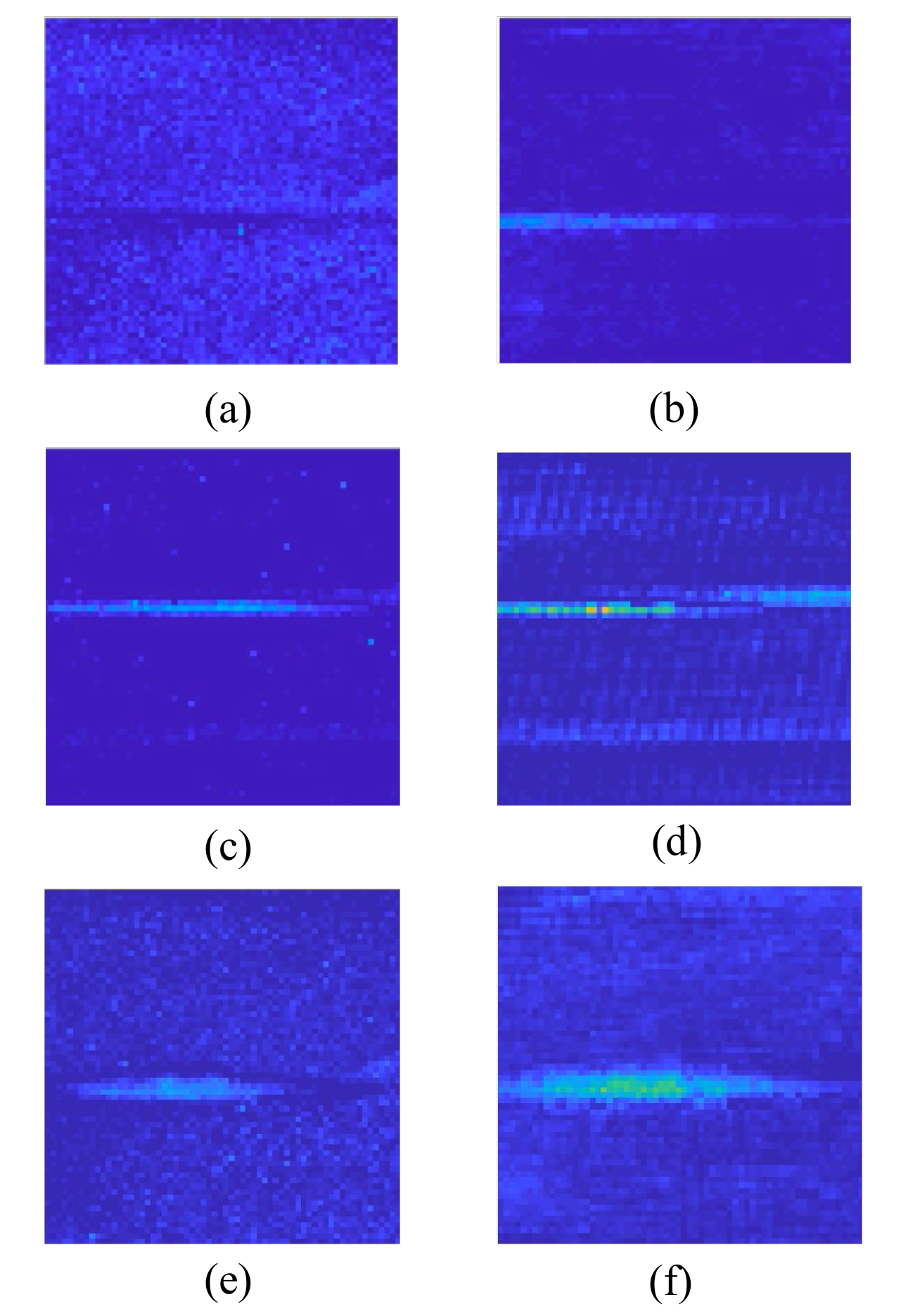}
    \caption{ACE detection of plumes for a number of chemical compounds. The left column is ACE detection on raw data, the right column is ACE detection on hyperspectral images that have undergone 10\% CS sampling followed by reconstruction via \eqref{eqn-hyper-opt-problem}. (a)-(b) and (e)-(f) are detection of MeS, while (c)-(d) are detection of the chemical TEP.}
    \label{fig_enhance_example}
\end{figure}

To give the reader a sense of what these improvements look like in practice, in Fig.~\ref{fig_enhance_example} we show three pairs of frames of ACE detection for both raw data (on the left) and data reconstructed from just $10\%$ CS sampling (on the right). One can see that there are certain instances (such as (a)-(b)) where CS sampling highlights a signal that otherwise would not be noticeable. In (c)-(d), we see that the chemical signal has been amplified but there is a question about whether it is as spatially accurate. Finally, in (e)-(f), CS sampling and reconstruction simply strengthens a signal that is already present, likely showing spatial regions where the signal was too weak to have been seen before. 

\subsection{Possible theoretical explanation and analysis}
\label{sect:explainations}

There are a number of possible explanations for the signal amplification phenomenon observed in hyperspectral data reconstruction by solving \eqref{eqn-opt-problem}. The first is that sampling $\tilde{U}$ and then using this sample to reconstruct the approximation $U$ of $\tilde{U}$ is effectively denoising $\tilde{U}$. Indeed, as was noted in Fig.~\ref{fig-TEPA_numberover_BCP}, the rate of false positives of TEP detection is strongly reduced in the reconstructed data compared to the raw data. On the other hand, the denoising effects are not supported by Fig.~\ref{fig-MESC_numberover} where high rates of false positives appear in both the raw and reconstructed data.



Another facet worth further investigation is the apparent trade-off between spatial accuracy and signal detection. As can be seen in Fig.~\ref{fig_enhance_example}, while signal strength is stronger in reconstructed cubes, the spatial accuracy may sometimes be reduced. This would be consistent with the sampling strategy used in this paper where sampling was done in the spatial domain but not the spectral domain. If, more generally, there is a tradeoff between spatial and spectral accuracy, then the phenomenon of signal amplification in reconstructed data will be more applicable to situations where detecting the presence of a chemical is valued above understanding precise spatial locations of chemicals within the field of view.


\section{Conclusion}
\label{sec:conclusion}

In this paper we described a phenomenon in which hyperspectral images sampled at very low levels and reconstructed using techniques from CS not only contain strong chemical signals, but sometimes even contain amplified signals. 
This observation suggests some new directions for research, the most important of which is to explain why this is happening. While we suggest some possible explanations in Section \ref{sect:explainations}, much more work needs to be done in this direction. Some other questions that should be investigated include:

\begin{itemize}
    \item Does this phenomenon occur when a different reconstruction framework (or optimization problem) is used? It would be especially interesting to know whether this occurs in some CS frameworks specifically designed for hyperspectral imaging (e.g. \cite{golbabaee2012hyperspectral}).
    \item Would similar results be obtained when using a different sparsifying basis (i.e. instead of the Haar wavelet basis)?
    \item In the experiments described in this paper, the same modified Walsh-Hadamard sampling basis was consistently used \cite{farnell2019sampling}. Does signal amplification still occur when different sampling strategies are used?
\end{itemize}

\acknowledgments 
The authors would like to thank Louis Scharf for insightful discussions related to this work, especially with regard to content involving ACE and MPACE. This research was partially supported by 
Department of Defense Army STTR Compressive Sensing Flash IR 3D Imager contract W911NF-16-C-0107
and Department of Energy STTR Compressive Spectral Video in the LWIR contract W911SR-17-C-0012.

\bibliography{report} 

\def\cprime{$'$} \def\cprime{$'$} \def\cprime{$'$} \def\cprime{$'$}
  \def\cprime{$'$} \def\cprime{$'$} \def\cprime{$'$}
\begin{thebibliography}{10}

\bibitem{martin2012new}
Martin, G., Dias, J. M.~B., and Plaza, A.~J., ``A new technique for
  hyperspectral compressive sensing using spectral unmixing,'' in [{\em
  Satellite Data Compression, Communications, and Processing
  VIII}{\nolinebreak\hspace{0.1em}]},   {\bf 8514},  85140N, International
  Society for Optics and Photonics (2012).

\bibitem{martin2013hyperspectral}
Martin, G., Bioucas-Dias, J.~M., and Plaza, A., ``Hyperspectral coded aperture
  ({HYCA}): A new technique for hyperspectral compressive sensing,'' in [{\em
  Signal Processing Conference (EUSIPCO), 2013 Proceedings of the 21st
  European}{\nolinebreak\hspace{0.1em}]},   1--5, IEEE (2013).

\bibitem{martin2015hyca}
Mart{\'\i}n, G., Bioucas-Dias, J.~M., and Plaza, A., ``Hyca: A new technique
  for hyperspectral compressive sensing,'' {\em IEEE Transactions on Geoscience
  and Remote Sensing}~{\bf 53}(5),  2819--2831 (2015).

\bibitem{golbabaee2012hyperspectral}
Golbabaee, M. and Vandergheynst, P., ``Hyperspectral image compressed sensing
  via low-rank and joint-sparse matrix recovery,'' in [{\em Acoustics, Speech
  and Signal Processing (ICASSP), 2012 IEEE International Conference
  on}{\nolinebreak\hspace{0.1em}]},   2741--2744, Ieee (2012).

\bibitem{rajwade2013coded}
Rajwade, A., Kittle, D., Tsai, T.-H., Brady, D., and Carin, L., ``Coded
  hyperspectral imaging and blind compressive sensing,'' {\em SIAM Journal on
  Imaging Sciences}~{\bf 6}(2),  782--812 (2013).

\bibitem{baraniuk2007compressive}
Baraniuk, R.~G., ``Compressive sensing [lecture notes],'' {\em IEEE signal
  processing magazine}~{\bf 24}(4),  118--121 (2007).

\bibitem{rudin1992nonlinear}
Rudin, L.~I., Osher, S., and Fatemi, E., ``Nonlinear total variation based
  noise removal algorithms,'' {\em Physica D: nonlinear phenomena}~{\bf
  60}(1-4),  259--268 (1992).

\bibitem{farnell2019sampling}
Farnell, E., Kvinge, H., Dixon, J.~P., Dupuis, J.~R., Kirby, M., Peterson, C.,
  Schundler, E.~C., and Smith, C.~W., ``A data-driven approach to sampling
  matrix selection for compressive sensing,'' (2019).
\newblock Preprint arXiv:1906.08869.

\bibitem{GO09}
Goldstein, T. and Osher, S., ``The split {B}regman method for
  {$L1$}-regularized problems,'' {\em SIAM J. Imaging Sci.}~{\bf 2}(2),
  323--343 (2009).

\bibitem{scharf1996adaptive}
Scharf, L.~L. and McWhorter, L.~T., ``Adaptive matched subspace detectors and
  adaptive coherence estimators,'' in [{\em Signals, Systems and Computers,
  1996. Conference Record of the Thirtieth Asilomar Conference
  on}{\nolinebreak\hspace{0.1em}]},   1114--1117, IEEE (1996).

\bibitem{kraut2001adaptive}
Kraut, S., Scharf, L.~L., and McWhorter, L.~T., ``Adaptive subspace
  detectors,'' {\em IEEE Transactions on signal processing}~{\bf 49}(1),  1--16
  (2001).

\bibitem{pakrooh2017adaptive}
Pakrooh, P., Scharf, L.~L., Cheney, M., Homan, A., and Ferrara, M.,
  ``Multipulse adaptive coherence for detection in wind turbine clutter,'' {\em
  IEEE Transactions on Aerospace and Electronic Systems}~{\bf 53}(6),
  3091--3103 (2017).

\bibitem{pakrooh2017adaptiveb}
Pakrooh, P., Scharf, L., Cheney, M., Homan, A., and Ferrara, M., ``The adaptive
  coherence estimator for detection in wind turbine clutter,'' in [{\em Radar
  Conference (RadarConf), 2017 IEEE}{\nolinebreak\hspace{0.1em}]},
  1793--1798, IEEE (2017).

\bibitem{scharf2017multipulse}
Scharf, L.~L. and Pakrooh, P., ``Multipulse subspace detectors,'' in [{\em
  Conference record-Asilomar Conference on Signals, Systems, \&
  Computers}{\nolinebreak\hspace{0.1em}]},  (2017).

\bibitem{farnell2019TVvsL1}
Farnell, E., Kvinge, H., Dupuis, J.~R., Kirby, M., Peterson, C., and Schundler,
  E.~C., ``Total variation vs {L}1 regularization: a comparison of compressive
  sensing optimization methods for chemical detection,'' (2019).
\newblock Preprint arXiv:1906.10603.

\bibitem{cosofret2009airis}
Cosofret, B.~R., Chang, S., Finson, M.~L., Gittins, C.~M., Janov, T.~E., Konno,
  D., Marinelli, W.~J., Levreault, M.~J., and Miyashiro, R.~K., ``{AIRIS}
  standoff multispectral sensor,'' in [{\em Chemical, Biological, Radiological,
  Nuclear, and Explosives (CBRNE) Sensing X}{\nolinebreak\hspace{0.1em}]},
  {\bf 7304},  73040Y, International Society for Optics and Photonics (2009).

\bibitem{broadwater2011primer}
Broadwater, J.~B., Limsui, D., and Carr, A.~K., ``A primer for chemical plume
  detection using {LWIR} sensors,'' {\em Technical Paper, National Security
  Technology Department, Las Vegas, NV}  (2011).

\end{thebibliography}
\bibliographystyle{spiebib} 

\end{document}